\documentclass[roman,11pt]{article}

\usepackage{amsmath,amssymb,amsthm,amscd,latexsym}
\usepackage{times}
\usepackage{hyperref}
\usepackage{cite}
\usepackage{w-greek,wasysym,graphicx}

\def\rfr#1{eq. (\ref{#1})}
\def\rfrs#1#2{eq. (\ref{#1})-eq. (\ref{#2})}

\def\virg#1{``#1''}
\def\eqi{\begin{equation}}
\def\eqf{\end{equation}}
\def\rp#1#2{{#1\over#2}} \def\lb#1{\label{#1}}
\def\bds#1{\boldsymbol{#1}}
\def\ton#1{\left(#1\right)}

\def\grf#1{\left\{#1\right\}}

\begin{document}

\title{Frame-Dragging in Extrasolar Circumbinary Planetary Systems}

\author{L. Iorio\\ Ministero dell'Istruzione, dell'Universit$\grave{\textrm{a}}$ e della Ricerca (M.I.U.R.)-Istruzione \\ Fellow of the Royal Astronomical Society (F.R.A.S.)\\ Viale Unit$\grave{\textrm{a}}$ di Italia 68, 70125, Bari (BA), Italy}

\maketitle

\begin{abstract}
Extrasolar circumbinary planets are so called because they orbit two stars instead of just one; to date, an increasing number of such planets have been discovered with a variety of techniques. If the orbital frequency of the hosting stellar pair is much higher than the planetary one, the tight stellar binary can be considered as a matter ring current generating its own post-Newtonian stationary gravitomagnetic field through its orbital angular momentum. It affects the orbital motion of a relatively distant planet with Lense-Thirring-type precessional effects which, under certain circumstances, may amount to a significant fraction of the static, gravitoelectric ones, analogous to the well known Einstein perihelion precession of Mercury, depending only on the masses of the system's bodies. Instead, when the gravitomagnetic field is due solely to the spin of each of the central star(s), the Lense-Thirring shifts are several orders of magnitude smaller than the gravitoelectric ones. In view of the growing interest in the scientific community about the detection of general relativistic effects in exoplanets, the perspectives of finding new scenarios for testing such a further manifestation of general relativity might be deemed worth of further investigations.
\end{abstract}



\centerline
{PACS: 04.80.-y; 04.80.Cc;  97.82.-j}

\section{Introduction}

According to the General Theory of Relativity\footnote{For a recent overview, see, e.g., \cite{2016Univ....2...23D} and references therein.} (GTR), the deformed spacetime generated by a localized, non-static distribution of matter-energy such as a rotating star affects the orbital motion of a nearby test particle like, e.g., a planet p in such a way that its trajectory is not closed, as in the case of the unchanging Keplerian ellipse of the Newtonian mechanics. Among other things, there are two types of resulting secular effects to the first post-Newtonian (1pN) order: the static \virg{gravitoelectric} (GE) Einstein precession \cite{Ein15} of the pericenter $\omega_\mathrm{p}$ due solely to the total mass $M$ of the system, and the stationary \virg{gravitomagnetic} Lense-Thirring (LT) precessions \cite{LT18} of the longitude of the ascending node $\Omega_\mathrm{p}$ and the pericenter  caused by the proper angular momentum $\bds J$ of the spinning central object; see, e.g., \cite{1991ercm.book.....B,SoffelHan19}. In general, the pN gravitomagnetic field is generated by mass-energy currents; for the concept of gravitoelectromagnetism within general relativity, see, e.g., \cite{1958NCim...10..318C,Thorne86,1986hmac.book..103T,1988nznf.conf..573T,1991AmJPh..59..421H,1992AnPhy.215....1J,2001rfg..conf..121M,2001rsgc.book.....R,Mash07,2008PhRvD..78b4021C,
2014GReGr..46.1792C,2021Univ....7..388C,2021Univ....7..451R}, and references therein. The gravitoelectric and gravitomagnetic net shifts per orbit of the planet's pericentre $\omega_\mathrm{p}$ are \cite{1975PhRvD..12..329B,1988NCimB.101..127D,2017EPJC...77..439I}
\begin{align}
\Delta\omega_\mathrm{p}^{\rm GE} \lb{oGE}&= \rp{6\,\uppi\,G\,M}{c^2\,a_\mathrm{p}\,\ton{1-e_\mathrm{p}^2}},\\ \nonumber \\
\Delta\omega_\mathrm{p}^{\rm LT} \lb{oLT} &= -\rp{4\,\uppi\,G\,\bds{J}\bds\cdot\ton{2\,{\bds{\hat{h}}}_\mathrm{p} + \cot I_\mathrm{p}\,{\bds{\hat{m}}}_\mathrm{p}}}{c^2\,n_\mathrm{p}\,a_\mathrm{p}^3\,\ton{1-e_\mathrm{p}^2}^{3/2}},
\end{align}
where $G$ is the Newtonian constant of gravitation, $c$ is the speed of light in vacuum, $a_\mathrm{p}$ is the planet's semimajor axis, $e_\mathrm{p}$ is the planet's eccentricity, $I_\mathrm{p}$ is the inclination of the planetary orbital plane to the reference $\grf{x,\,y}$ plane, customarily identified with the plane of the sky,
\eqi
n_{\rm p}\doteq\sqrt{\rp{G\,M}{a_\mathrm{p}^3}}\lb{Kep}
\eqf
is the planet's Keplerian mean motion,
\eqi
{\bds{\hat{h}}}_\mathrm{p} = \grf{\sin I_\mathrm{p}\,\sin\Omega_\mathrm{p},\,-\sin I_\mathrm{p}\,\cos\Omega_\mathrm{p},\,\cos I_\mathrm{p}}\lb{enne}
\eqf
is a unit vector directed along the planet's orbital angular momentum, and
\eqi
{\bds{\hat{m}}}_\mathrm{p} = \grf{-\cos I_\mathrm{p}\,\sin\Omega_\mathrm{p},\,\cos I_\mathrm{p}\,\cos\Omega_\mathrm{p},\,\sin I_\mathrm{p}}\lb{emme}
\eqf
is a unit vector in the planetary orbital plane perpendicular to the line of the nodes, which is the intersection of the orbital plane with the reference $\grf{x,y}$ plane \cite{1991ercm.book.....B,SoffelHan19}.
For the sake of completeness, also the Lense-Thirring node precession is mentioned: its net shift per orbit is \cite{1975PhRvD..12..329B,2017EPJC...77..439I}
\eqi
\Delta\Omega_\mathrm{p}^{\rm LT} \lb{OLT} = \rp{4\,\uppi\,G\,\csc I_\mathrm{p}\,\bds{J}\bds\cdot\bds{\hat{m}}_\mathrm{p}}{c^2\,n_\mathrm{p}\,a_\mathrm{p}^3\,\ton{1-e_\mathrm{p}^2}^{3/2}}.
\eqf

As it turns out from \rfrs{oGE}{oLT}, the gravitoelectric effect is, in general, quite larger than the gravitomagnetic one; suffice it to say that in the case of Sun and Mercury\footnote{In \rfrs{merc}{sats}, \rfr{oLT} is computed in a coordinate system whose reference $\grf{x,\,y}$ plane is aligned with the primary's equatorial plane, i.e., $\bds{\hat{J}} = \grf{0,\,0,\,1}$.} it is
\eqi
\left|\rp{\Delta\omega_{\mercury}^{\rm LT}}{\Delta\omega_{\mercury}^{\rm GE}}\right|\approx 3\times 10^{-5},\lb{merc}
\eqf
while for the Earth and, say, the LAGEOS (L) satellite \cite{Lucchesi019}, it is
\eqi
\left|\rp{\Delta\omega_\mathrm{L}^{\rm LT}}{\Delta\omega_\mathrm{L}^{\rm GE}}\right|\approx 9\times 10^{-3}.\lb{sats}
\eqf
This is why the gravitoelectric precessions have been known since the observations of the then anomalous Mercury's motion by Le Verrier \cite{LeVerrier1859} in the mid-nineteenth century, while the Lense-Thirring effect is still so difficult to measure with both natural and artificial objects \cite{2013CEJPh..11..531R,Everittetal011}.

Nevertheless, for  planets orbiting a binary star  \cite{2010ASSL..366.....H,2015pes..book..309T}, known as circumbinary planets (CBPs), some of which have already been discovered with different techniques \cite{1993ApJ...412L..33T,2005A&A...440..751C,2009AJ....137.3181L,2010ApJ...708L..66Q,2010A&A...521L..60B,2011Sci...333.1602D,2012ApJ...758...87O,2012Sci...337.1511O,
2012ApJ...745L..23Q,2012MNRAS.422L..24Q,2012Natur.481..475W,2013ApJ...768..127S,2014ApJ...781...20K,
2014ApJ...784...14K,2015ApJ...809...26W,2016AJ....152..125B,2016ApJ...827...86K,2017MNRAS.468.2932G,2017MNRAS.468L.118J,2018A&A...619A..43A,2021AJ....162..234K}, certain gravitomagnetic effects may be much larger than expected, amounting to about $10\%$  or so of the gravitoelectric ones.
The basic idea is as follows. The CBPs discovered so far can be considered as hierarchical triple systems consisting of an inner binary star b and a distant planet p that orbits the centre of mass of b. Thus,  the two inner stars A and B can be approximately considered as  a mass current sourcing a gravitomagnetic field much stronger than that due to the individual spins of each star through the binary's orbital angular momentum
\eqi
J_\mathrm{b} =\mu_\mathrm{b}\,\sqrt{G\,M_\mathrm{b}\,a_\mathrm{b}\,\ton{1-e^2_\mathrm{b}}},\lb{Lb}
\eqf
where
\eqi
\mu_\mathrm{b} \doteq \rp{M_\mathrm{A}\,M_\mathrm{B}}{M_\mathrm{b}}
\eqf
is the binary's reduced mass, $M_\mathrm{b} \doteq M_\mathrm{A}+M_\mathrm{B}$ is the total mass of the binary, $a_\mathrm{b}$ is the binary's semimajor axis and $e_\mathrm{b}$ is the binary's eccentricity.
Thus, it is expected that \rfr{oLT}, calculated with\footnote{In \rfrs{oGE}{Kep}, $M$ is now meant as $M_\mathrm{b} + M_\mathrm{p}$, where $M_\mathrm{p}$ is the planet's mass.} \rfr{Lb}, yields a much larger pericenter precession.

Testing relativistic frame dragging in as much ways as possible is important since it is believed to play important roles in several high-energy astrophysical phenomena in strong-field systems \cite{1975ApJ...195L..65B,Thorne86,1988nznf.conf..573T,1998ApJ...492L..59S,Pen2002,2009SSRv..148...37S,StellaPossenti09}. Extending  relativistic gravitomagnetism with confidence to such relatively unknown scenarios, for which no direct access is available, requires that it is corroborated in more than just a single case \cite{Everittetal011}.

\section{The gravitomagnetic precessions due to a matter ring current}

\begin{figure}[ht!]
\centering
\begin{tabular}{cc}
\includegraphics[width = 6 cm]{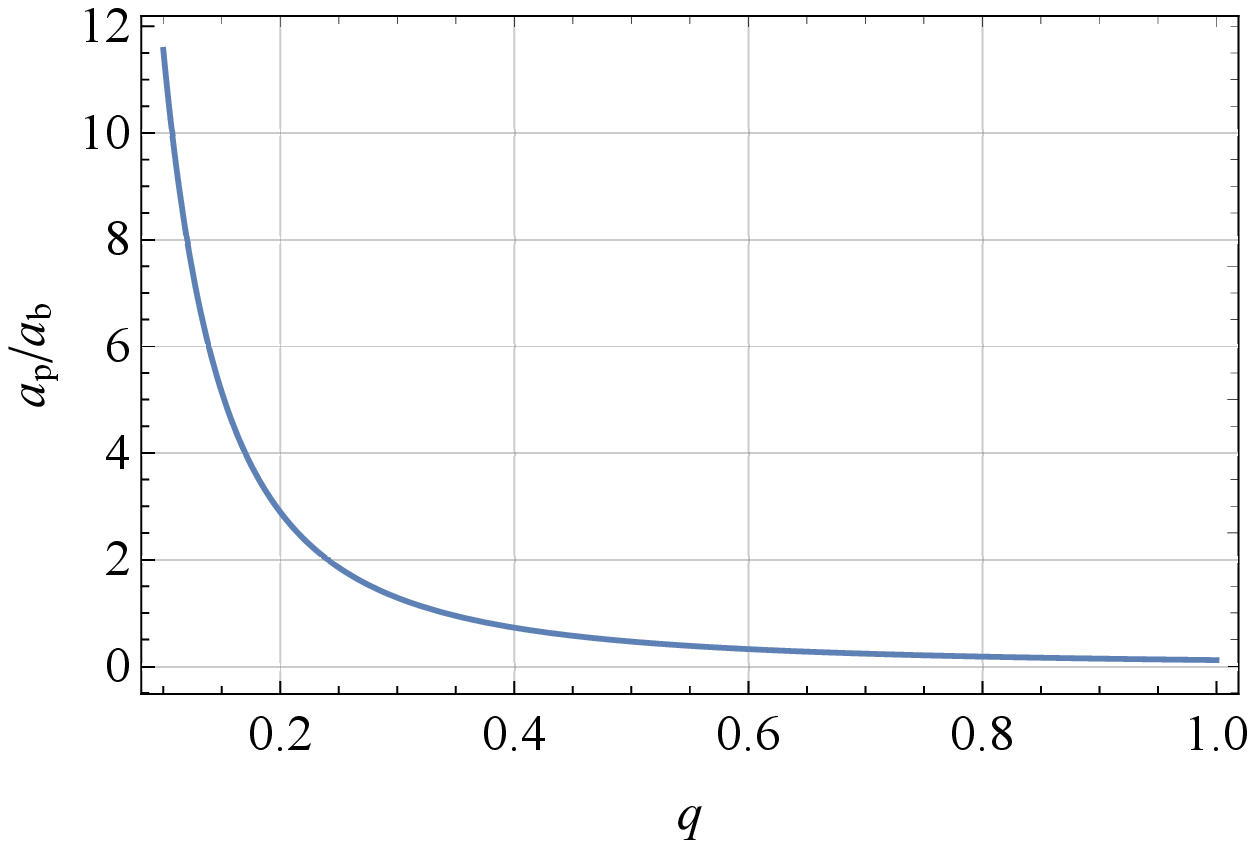} & \includegraphics[width = 6 cm]{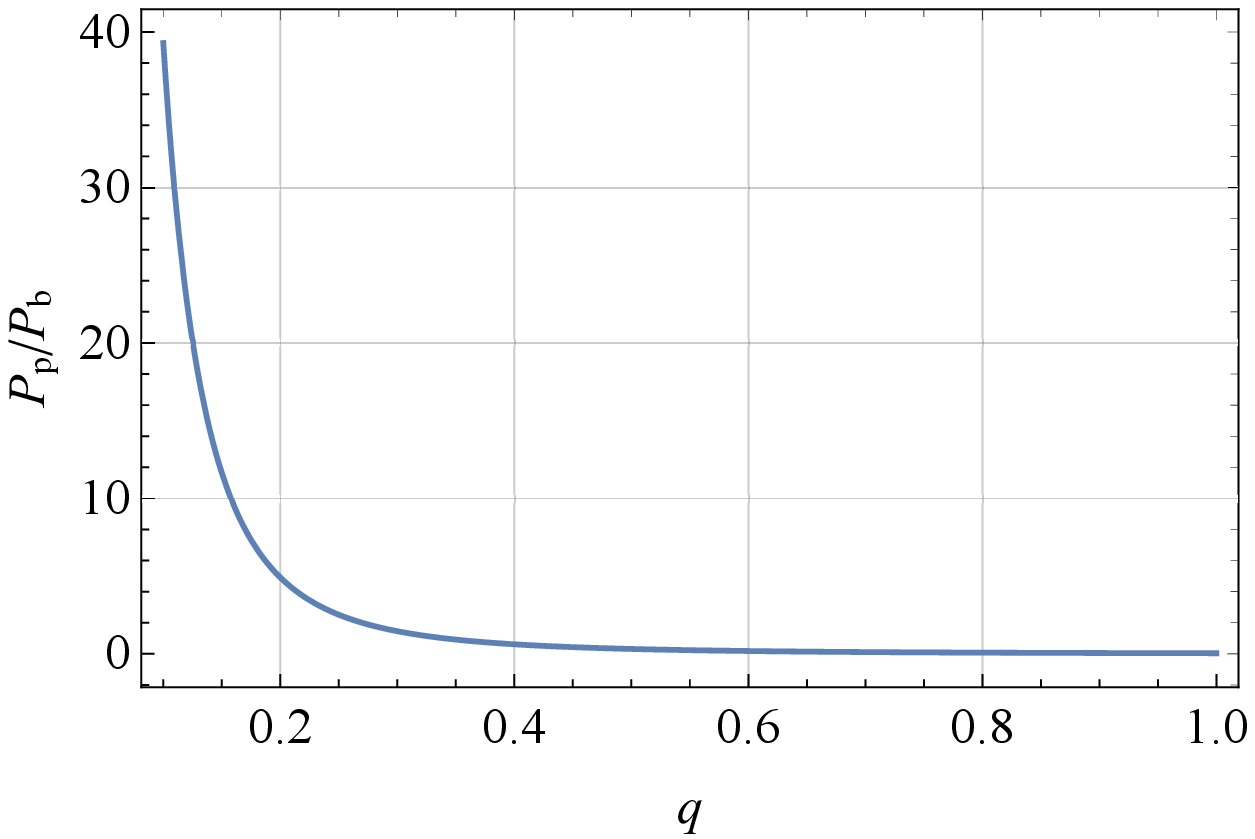}\\
\end{tabular}
\caption{
Left panel: ratio of the planet's semimajor axis $a_\mathrm{p}$ to the inner binary's one $a_\mathrm{b}$ as a function of $q$ according to \rfr{qu}. Right panel: same for the orbital periods of the planet ($P_\mathrm{p}$) and of the inner binary ($P_\mathrm{b}$).
}\label{fig0}
\end{figure}

By imposing
\eqi
\Delta\omega_\mathrm{p}^\mathrm{LT} = q\,\Delta\omega_\mathrm{p}^\mathrm{GE},
\eqf
with $q > 0$, yields the following condition for the semimajor axis of the planet's orbit about the inner binary
\eqi
a_\mathrm{p} = \rp{16\,a_\mathrm{b}\,\ton{1-e_\mathrm{b}^2}\,M_\mathrm{A}^2\,M_\mathrm{B}^2}{9\,\ton{1-e_\mathrm{p}^2}\,M_\mathrm{b}\,\ton{M_\mathrm{b} + M_\mathrm{p}}^3\,q^2}.\lb{qu}
\eqf
As an example, for a binary of two Sun-like stars ($M_\mathrm{A} = M_\mathrm{B} = M_\odot$) in a circular orbit ($e_\mathrm{b}=0.0$) and a Jupiter-type ($M_\mathrm{p} = M_{\jupiter}$) circumbinary planet in a moderately eccentric orbit ($e_\mathrm{p}=0.2$), by imposing
\eqi
q = 0.1
\eqf
one gets
\eqi
a_\mathrm{p} = 11.5\,a_\mathrm{b},
\eqf
which fulfils the assumption that the binary is viewed by the planet as a rotating matter ring.
By setting
\eqi
q=1
\eqf
in \rfr{qu}, corresponding to
\eqi
\Delta\omega_\mathrm{p}^\mathrm{LT} = \Delta\omega_\mathrm{p}^\mathrm{GE},
\eqf
yields
\eqi
a_\mathrm{p} = 0.1\,a_\mathrm{b},
\eqf
which implies that the gravitomagnetic precession cannot be as large as the gravitoelectric one. From Figure\,\ref{fig0}, it turns out that $a_\mathrm{p} = a_\mathrm{b}$, and $P_\mathrm{p} = P_\mathrm{b}$,  for $q\simeq 0.34$. Thus, for CBPs, this form of Lense-Thirring effect cannot reach the $\simeq 30\%$ of the gravitoelectric one.

\begin{figure}[ht!]
\centering
\begin{tabular}{cc}
\includegraphics[width = 6 cm]{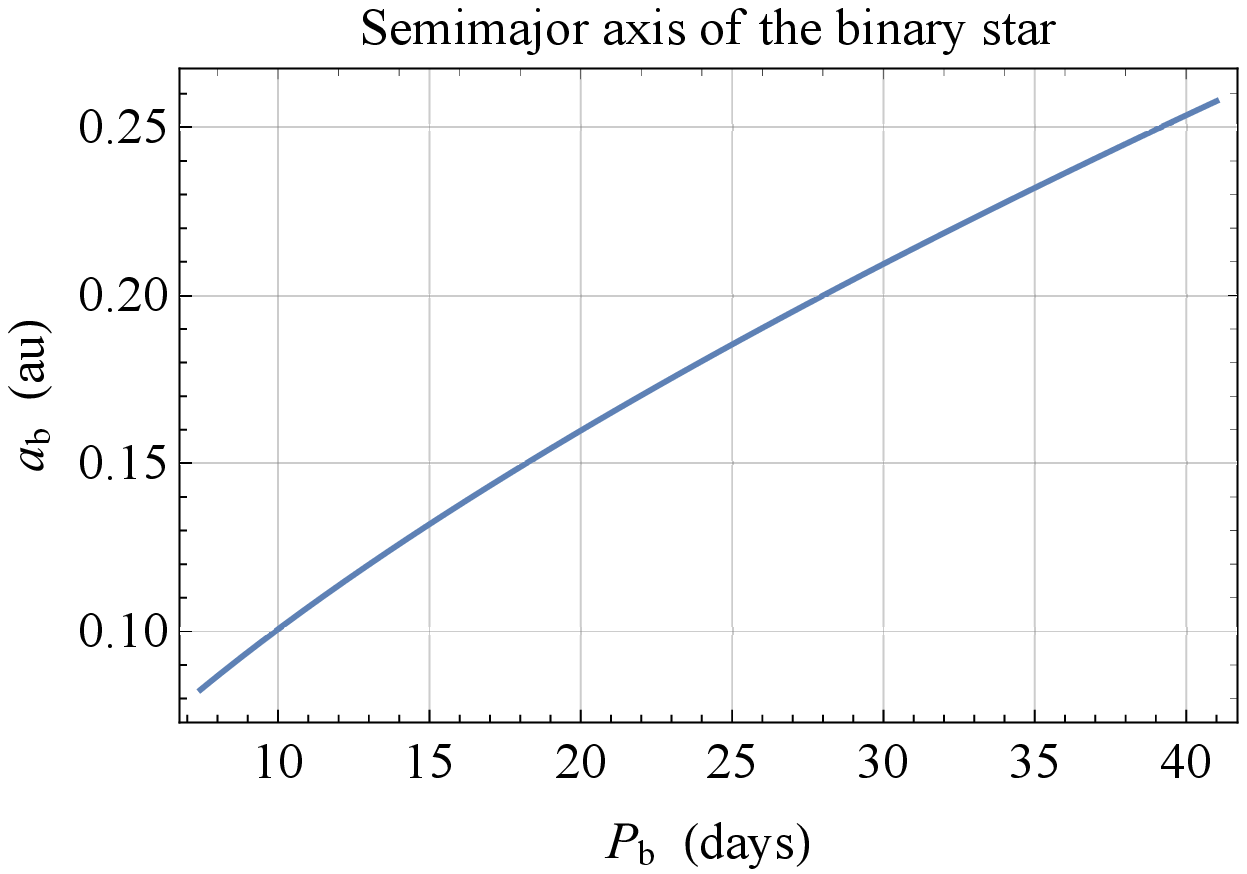} & \includegraphics[width = 6 cm]{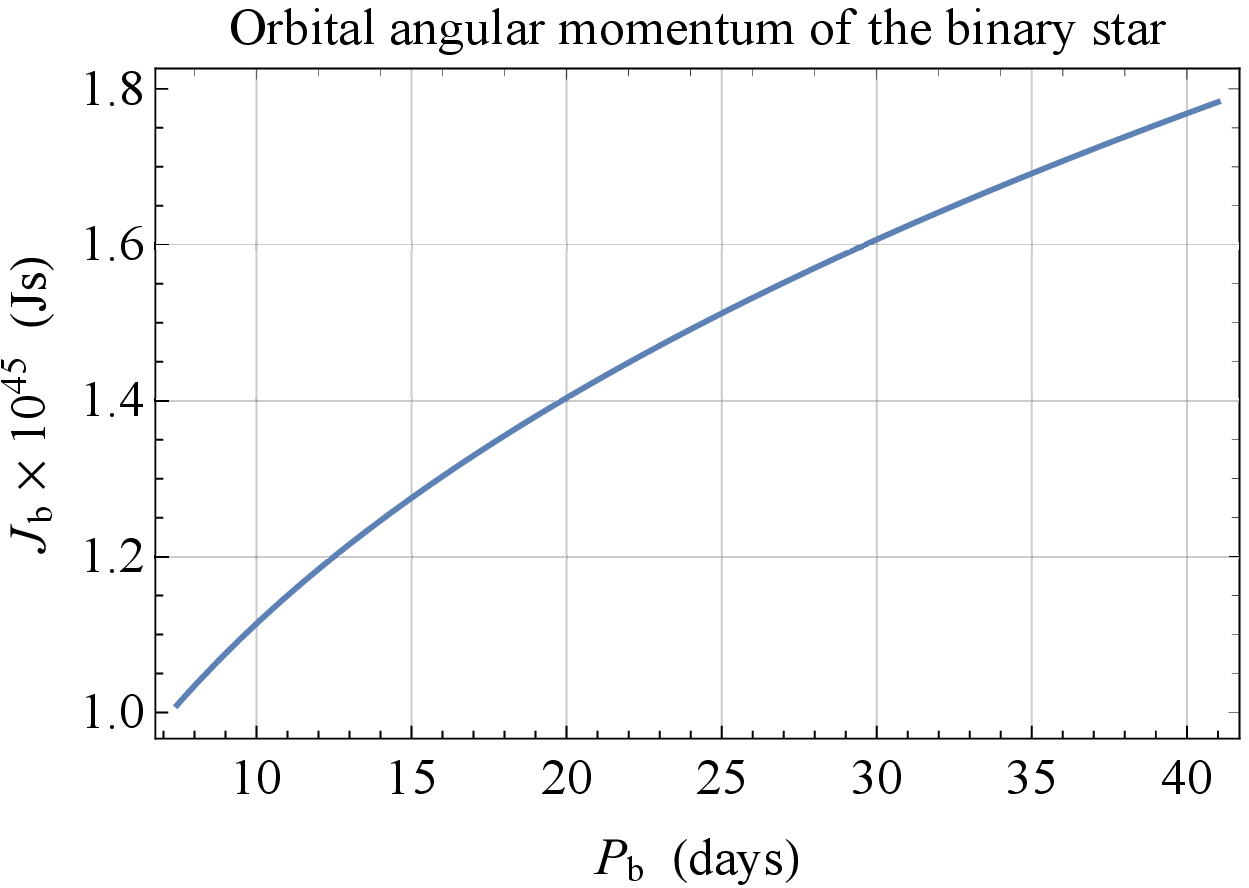}\\
\includegraphics[width = 6 cm]{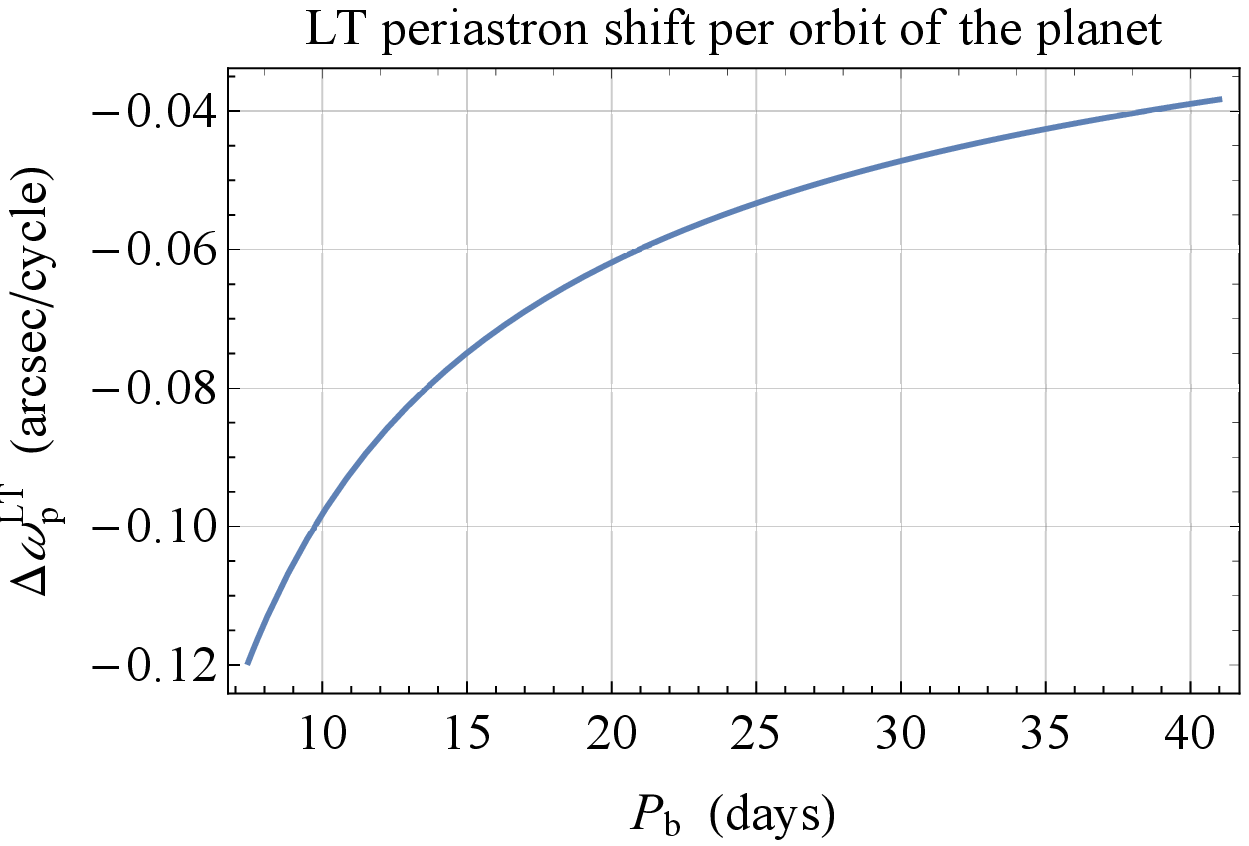} & \includegraphics[width = 6 cm]{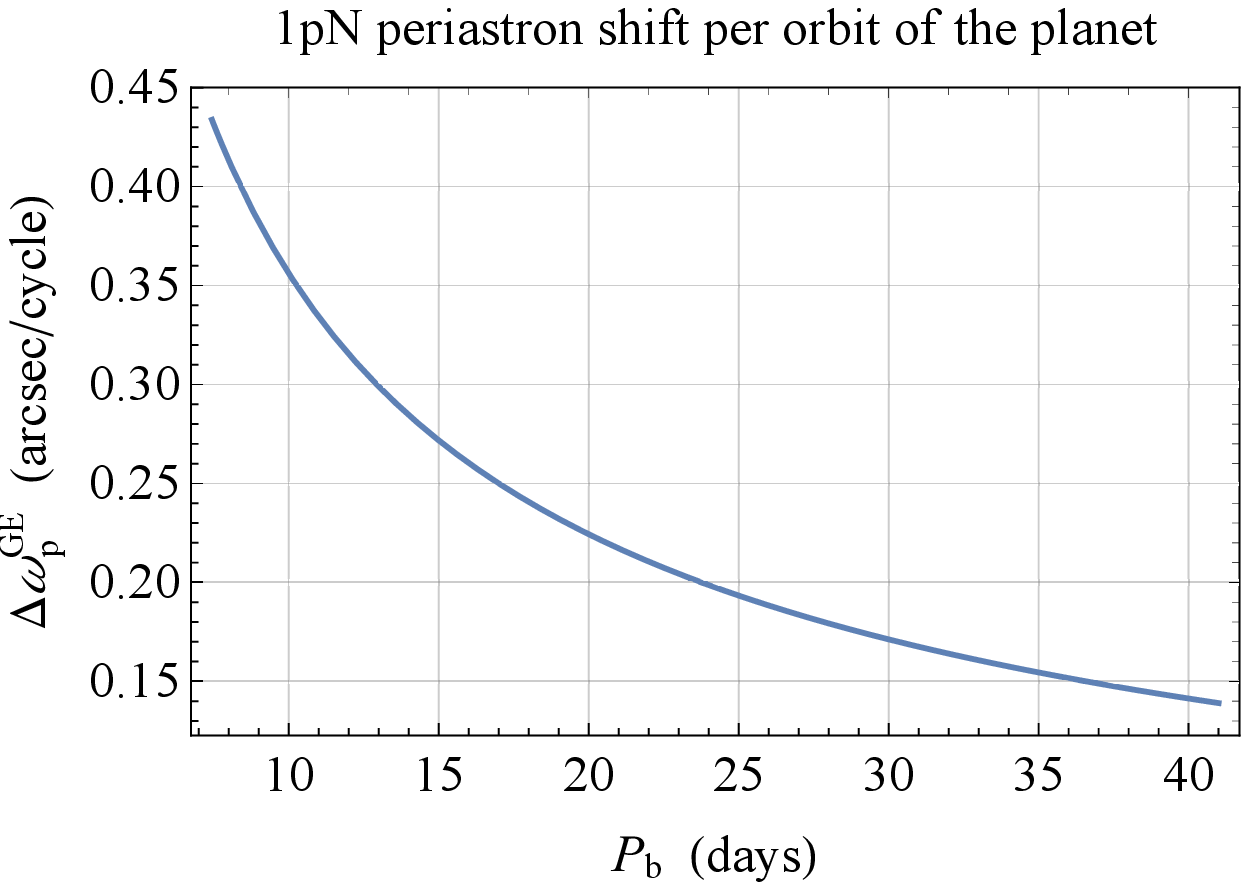}\\
\end{tabular}
\caption{
Upper row: binary's semimajor axis $a_\mathrm{b}$, in au, and orbital angular momentum $J_\mathrm{b}$, in J\,s, as functions of the binary's orbital period $P_\mathrm{b}$ ranging from $7.44$ to $41\,\mathrm{d}$.
Lower row: planet's gravitomagnetic and gravitoelectric net shifts per orbit, in $\mathrm{arcsec\,cycle}^{-1}$, as functions of the binary's orbital period $P_\mathrm{b}$ ranging from $7.44$ to $41\,\mathrm{d}$. The orbital stability condition $a_\mathrm{p}=2\,a_\mathrm{b}$ was adopted along with $M_\mathrm{A} = 0.69\,M_\odot = 1.03\,M_\mathrm{B},\,e_\mathrm{b} = 0.023,\,M_\mathrm{p} =  M_{\jupiter},\,e_\mathrm{p}=0.521$.
}\label{fig1}
\end{figure}

According to \cite{2014IAUS..293..125W}, most of the CBPs exhibit a high degree of coplanarity with the inner binary, i.e. ${\bds J}_\mathrm{b}$ and ${\bds{h}}_\mathrm{p}$ are almost aligned. Thus, the pericentre change of \rfr{oLT} can be approximated by
\eqi
\Delta\omega_\mathrm{p}^{\rm LT} \lb{oLT2} \simeq -\rp{8\,\uppi\,G\,J_\mathrm{b}}{c^2\,n_\mathrm{p}\,a_\mathrm{p}^3\,\ton{1-e_\mathrm{p}^2}^{3/2}},
\eqf
while the node shift of \rfr{OLT} almost vanishes.
The precession of \rfr{oLT2} is always negative, i.e. the pericentre moves in the opposite direction of the motion of the iner binary. There are some special cases, reported in  the literature, where negative orbital plane precessions were reported, e.g. in the presence of a Kerr naked singularity \cite{2017PhRvD..95h4024C} and of a hypothetical gravitomagnetic monopole \cite{2018PhRvD..98d3021C}.
Furthermore, \cite{2014IAUS..293..125W} remark that the mass of the primary star varies from $0.69$ to $1.53\,M_\odot$, with a mass ratio between $1.03$ and $3.76$ and eccentricity $0.023 \leq e_\mathrm{b}\leq 0.521$. As far as the CMPs are concerned, their orbital periods are in the range $7.44\,\mathrm{d}\leq P_\mathrm{p}\leq 41\,\mathrm{d}$, with eccentricities $e_\mathrm{p}$ varying from $0.007$ to $0.182$ \cite{2014IAUS..293..125W}.
In order to be stable around the binary host, the planet's orbit must be characterized by $a_\mathrm{p}=2-4\,a_\mathrm{b}$, a condition that is fulfilled by the CBPs considered  in \cite{2014IAUS..293..125W}. However, from the point of view of a possible detection of the sought effect, the issue of the stability of a discovered CBP is not relevant since its lifetime, even if short in astronomical terms, is certainly much longer than any conceivable  time span during which observations are collected.Figure\,\ref{fig1} displays $\Delta\omega_\mathrm{p}^\mathrm{LT}$ and $\Delta\omega_\mathrm{p}^\mathrm{GE}$, in $\mathrm{arcsec\,cycle}^{-1}$, as functions of the binary's orbital period $P_\mathrm{b}$, in d, by imposing the orbital stability condition $a_\mathrm{p} = 2\,a_\mathrm{b}$. For the remaining physical and orbital parameters, the values $M_\mathrm{A} = 0.69\,M_\odot = 1.03\,M_\mathrm{B},\,e_\mathrm{b} = 0.023,\,M_\mathrm{p} =  M_{\jupiter},\,e_\mathrm{p}=0.521$ were adopted. Among other things, also the binary's semimajor axis $a_\mathrm{b}$ and orbital angular momentum $J_\mathrm{b}$ are shown; it can be noted that $J_\mathrm{b}$ is about $10^4$ times larger than the spin angular momentum of the Sun which is of the order of $J_\odot\simeq 10^{41}\,\mathrm{J\,s}$ \cite{1998MNRAS.297L..76P}.
The size of the Lense-Thirring shift ranges from $0.12\,\mathrm{to}\,0.04\,\mathrm{arcsec\,cycle}^{-1}$, while the gravitoelectric one is in the range $0.45-0.15\,\mathrm{arcsec\,cycle}^{-1}$.

\section{Conclusions}

So far, a handful of circumbinary planets orbiting different types of stellar pairs, including compact objects as well, have been discovered;
although for all of them the matter ring current approximation is substantially valid for their hosting stellar pairs, they are likely too distant from them to allow for a measurement of relativistic effects.
Nonetheless, there may be reasons for being somewhat optimistic.

From the one hand, it is not unrealistic to expect that in a not too far future one or more systems with the right characteristics will be at our disposal.

On the other hand, there is a growing interest in the community of extrasolar planetary scientists about the possibility of extracting general relativistic signatures in such scenarios as well \cite{2006NewA...11..490I,2008ApJ...685..543J,2008MNRAS.389..191P,2009IAUS..253..492J,2009ApJ...698.1778R,2011A&A...535A.116D,2011PASJ...63..287F,2011Ap&SS.331..485I,
2011MNRAS.411..167I,2012MNRAS.423.1381E,2012ApJ...757..105K,2012Ap&SS.341..323L,2013RAA....13.1231Z,2019A&A...628A..80B,2021MNRAS.505.1567A,2021E&ES..658a2051G}.

\bibliography{circumbib}{}

\begin{thebibliography}{10}

\bibitem{2016Univ....2...23D}
I.~{Debono} and G.~F. {Smoot}.
\newblock {General Relativity and Cosmology: Unsolved Questions and Future
  Directions}.
\newblock {\em Universe}, 2:23, September 2016.

\bibitem{Ein15}
A.~Einstein.
\newblock Erkl\"{a}rung der perihelbewegung des merkur aus der allgemeinen
  relativit\"{a}tstheorie.
\newblock {\em Sitzungsberichte der Preu{\ss}ischen Akademie der
  Wissenschaften}, 47:831--839, November 1915.

\bibitem{LT18}
J.~Lense and H.~Thirring.
\newblock \"{U}ber den einflu{\ss} der eigenrotation der zentralk\"{o}rper auf
  die bewegung der planeten und monde nach der einsteinschen
  gravitationstheorie.
\newblock {\em Physikalische Zeitschrift}, 19:156--163, 1918.

\bibitem{1991ercm.book.....B}
V.~A. {Brumberg}.
\newblock {\em {Essential Relativistic Celestial Mechanics}}.
\newblock Adam Hilger, Bristol, 1991.

\bibitem{SoffelHan19}
M.~H. {Soffel} and W.-B. {Han}.
\newblock {\em {Applied General Relativity}}.
\newblock {Astronomy and Astrophysics Library}. Springer Nature Switzerland,
  Cham, 2019.

\bibitem{1958NCim...10..318C}
C.~{Cattaneo}.
\newblock {General relativity: Relative standard mass, momentum, energy and
  gravitational field in a general system of reference}.
\newblock {\em Il Nuovo Cimento}, 10(2):318--337, October 1958.

\bibitem{Thorne86}
K.~S. {Thorne}, D.~A. {MacDonald}, and R.~H. {Price}, editors.
\newblock {\em {Black Holes: The Membrane Paradigm}}.
\newblock Yale University Press, Yale, 1986.

\bibitem{1986hmac.book..103T}
K.~S. {Thorne}.
\newblock {Black Holes: The Membrane Viewpoint}.
\newblock In S.~L. {Shapiro}, S.~A. {Teukolsky}, and E.~E. {Salpeter}, editors,
  {\em Highlights of Modern Astrophysics: Concepts and Controversies}, pages
  103--161. Wiley, NY, 1986.

\bibitem{1988nznf.conf..573T}
K.~S. {Thorne}.
\newblock {Gravitomagnetism, jets in quasars, and the Stanford Gyroscope
  Experiment.}
\newblock In J.~D. {Fairbank}, Jr. {Deaver}, B.~S., C.~W.~F. {Everitt}, and
  P.~F. {Michelson}, editors, {\em Near Zero: New Frontiers of Physics}, pages
  573--586. Freeman, New York, 1988.

\bibitem{1991AmJPh..59..421H}
Edward~G. {Harris}.
\newblock {Analogy between general relativity and electromagnetism for slowly
  moving particles in weak gravitational fields}.
\newblock {\em American Journal of Physics}, 59(5):421--425, May 1991.

\bibitem{1992AnPhy.215....1J}
Robert~T. {Jantzen}, Paolo {Carini}, and Donato {Bini}.
\newblock {The many faces of gravitoelectromagnetism}.
\newblock {\em Annals of Physics}, 215(1):1--50, April 1992.

\bibitem{2001rfg..conf..121M}
B.~{Mashhoon}.
\newblock {Gravitoelectromagnetism}.
\newblock In J.~F. {Pascual-S{\'a}nchez}, L.~{Flor{\'\i}a}, A.~{San Miguel},
  and F.~{Vicente}, editors, {\em Reference Frames and Gravitomagnetism}, pages
  121--132. World Scientific, Singapore, July 2001.

\bibitem{2001rsgc.book.....R}
W.~{Rindler}.
\newblock {\em {Relativity: special, general, and cosmological}}.
\newblock Oxford University Press, Oxford, 2001.

\bibitem{Mash07}
B.~{Mashhoon}.
\newblock {Gravitoelectromagnetism: A Brief Review}.
\newblock In L.~{Iorio}, editor, {\em The Measurement of Gravitomagnetism: A
  Challenging Enterprise}, pages 29--39. Nova Science, New York, 2007.

\bibitem{2008PhRvD..78b4021C}
L.~Filipe~O. {Costa} and Carlos A.~R. {Herdeiro}.
\newblock {Gravitoelectromagnetic analogy based on tidal tensors}.
\newblock {\em Physical Review D}, 78(2):024021, July 2008.

\bibitem{2014GReGr..46.1792C}
L.~Filipe~O. {Costa} and Jos{\'e} {Nat{\'a}rio}.
\newblock {Gravito-electromagnetic analogies}.
\newblock {\em General Relativity and Gravitation}, 46:1792, October 2014.

\bibitem{2021Univ....7..388C}
L.~Filipe.~O. {Costa} and Jos{\'e} {Nat{\'a}rio}.
\newblock {Frame-Dragging: Meaning, Myths, and Misconceptions}.
\newblock {\em Universe}, 7(10):388, October 2021.

\bibitem{2021Univ....7..451R}
Matteo~Luca {Ruggiero}.
\newblock {A Note on the Gravitoelectromagnetic Analogy}.
\newblock {\em Universe}, 7(11):451, November 2021.

\bibitem{1975PhRvD..12..329B}
B.~M. {Barker} and R.~F. {O'Connell}.
\newblock {Gravitational two-body problem with arbitrary masses, spins, and
  quadrupole moments}.
\newblock {\em Physical Review D}, 12(2):329--335, July 1975.

\bibitem{1988NCimB.101..127D}
T.~{Damour} and G.~{Sch\"{a}fer}.
\newblock {Higher-order relativistic periastron advances and binary pulsars.}
\newblock {\em Il Nuovo Cimento B}, 101:127--176, February 1988.

\bibitem{2017EPJC...77..439I}
L.~{Iorio}.
\newblock {Post-Keplerian perturbations of the orbital time shift in binary
  pulsars: an analytical formulation with applications to the galactic center}.
\newblock {\em European Physics Journal C}, 77(7):439, July 2017.

\bibitem{Lucchesi019}
D.~M. {Lucchesi}, L.~{Anselmo}, M.~{Bassan}, C.~{Magnafico}, C.~{Pardini},
  R.~{Peron}, G.~{Pucacco}, and M.~{Visco}.
\newblock {General Relativity measurements in the field of Earth with
  laser-ranged satellites: state of the art and perspectives}.
\newblock {\em Universe}, 5(6):141, June 2019.

\bibitem{LeVerrier1859}
U.-J. Le~Verrier.
\newblock Lettre de {M.} {Le} {Verrier} \`{a} {M.} {Faye} sur la th\'{e}orie de
  {Mercure} et sur le mouvement du p\'{e}rih\'{e}lie de cette plan\`{e}te.
\newblock {\em Comptes rendus hebdomadaires des s\'{e}ances de l'Acad\'{e}mie
  des sciences}, 49:379--383, Juillet-D\'{e}cembre 1959.

\bibitem{2013CEJPh..11..531R}
G.~{Renzetti}.
\newblock {History of the attempts to measure orbital frame--dragging with
  artificial satellites}.
\newblock {\em Central European Journal of Physics}, 11(5):531--544, May 2013.

\bibitem{Everittetal011}
C.~W.~F. Everitt, D.~B. Debra, B.~W. Parkinson, J.~P. Turneaure, J.~W. Conklin,
  M.~I. Heifetz, G.~M. Keiser, A.~S. Silbergleit, T.~Holmes, J.~Kolodziejczak,
  M.~Al-Meshari, J.~C. Mester, B.~Muhlfelder, V.~G. Solomonik, K.~Stahl, Jr.
  Worden, P.~W., W.~Bencze, S.~Buchman, B.~Clarke, A.~Al-Jadaan, H.~Al-Jibreen,
  J.~Li, J.~A. Lipa, J.~M. Lockhart, B.~Al-Suwaidan, M.~Taber, and S.~Wang.
\newblock Gravity probe b: Final results of a space experiment to test general
  relativity.
\newblock {\em Physical Review Letters}, 106(22):221101, June 2011.

\bibitem{2010ASSL..366.....H}
N.~{Haghighipour}, editor.
\newblock {\em {Planets in Binary Star Systems}}, volume 366 of {\em
  Astrophysics and Space Science Library}. Springer, Berlin, 2010.

\bibitem{2015pes..book..309T}
P.~{Thebault} and N.~{Haghighipour}.
\newblock {Planet Formation in Binaries}.
\newblock In S.~{Jin}, N.~{Haghighipour}, and W.H. {Ip}, editors, {\em
  Planetary Exploration and Science: Recent Results and Advances}, pages
  309--340. Springer, Heidelberg, 2015.

\bibitem{1993ApJ...412L..33T}
S.~E. {Thorsett}, Z.~{Arzoumanian}, and J.~H. {Taylor}.
\newblock {PSR B1620-26 - A binary radio pulsar with a planetary companion?}
\newblock {\em The Astrophysical Journal Letters}, 412(1):L33--L36, July 1993.

\bibitem{2005A&A...440..751C}
A.~C.~M. {Correia}, S.~{Udry}, M.~{Mayor}, J.~{Laskar}, D.~{Naef}, F.~{Pepe},
  D.~{Queloz}, and N.~C. {Santos}.
\newblock {The CORALIE survey for southern extra-solar planets. XIII. A pair of
  planets around HD{\,}202206 or a circumbinary planet?}
\newblock {\em Astronomy and Astrophysics}, 440(2):751--758, September 2005.

\bibitem{2009AJ....137.3181L}
J.~W. {Lee}, S.-L. {Kim}, C.-H. {Kim}, R.~H. {Koch}, C.-U. {Lee}, H.-I. {Kim},
  and J.-H. {Park}.
\newblock {The sdB+M Eclipsing System HW Virginis and its Circumbinary
  Planets}.
\newblock {\em The Astromomical Journal}, 137(2):3181--3190, February 2009.

\bibitem{2010ApJ...708L..66Q}
S.-B. {Qian}, W.-P. {Liao}, L.-Y. {Zhu}, and Z.-B. {Dai}.
\newblock {Detection of a Giant Extrasolar Planet Orbiting the Eclipsing Polar
  DP Leo}.
\newblock {\em The Astrophysical Journal Letters}, 708(1):L66--L68, January
  2010.

\bibitem{2010A&A...521L..60B}
K.~{Beuermann}, F.~V. {Hessman}, S.~{Dreizler}, T.~R. {Marsh}, S.~G. {Parsons},
  D.~E. {Winget}, G.~F. {Miller}, M.~R. {Schreiber}, W.~{Kley}, V.~S.
  {Dhillon}, S.~P. {Littlefair}, C.~M. {Copperwheat}, and J.~J. {Hermes}.
\newblock {Two planets orbiting the recently formed post-common envelope binary
  NN Serpentis}.
\newblock {\em Astronomy and Astrophysics}, 521:L60, October 2010.

\bibitem{2011Sci...333.1602D}
L.~R. {Doyle}, J.~A. {Carter}, D.~C. {Fabrycky}, R.~W. {Slawson}, S.~B.
  {Howell}, J.~N. {Winn}, J.~A. {Orosz}, A.~{Pr\v{\ }sa}, W.~F. {Welsh}, S.~N.
  {Quinn}, D.~{Latham}, G.~{Torres}, L.~A. {Buchhave}, G.~W. {Marcy}, J.~J.
  {Fortney}, A.~{Shporer}, E.~B. {Ford}, J.~J. {Lissauer}, D.~{Ragozzine},
  M.~{Rucker}, N.~{Batalha}, J.~M. {Jenkins}, W.~J. {Borucki}, D.~{Koch}, C.~K.
  {Middour}, J.~R. {Hall}, S.~{McCauliff}, M.~N. {Fanelli}, E.~V. {Quintana},
  M.~J. {Holman}, D.~A. {Caldwell}, M.~{Still}, R.~P. {Stefanik}, W.~R.
  {Brown}, G.~A. {Esquerdo}, S.~{Tang}, G.~{Furesz}, J.~C. {Geary},
  P.~{Berlind}, M.~L. {Calkins}, D.~R. {Short}, J.~H. {Steffen}, D.~{Sasselov},
  E.~W. {Dunham}, W.~D. {Cochran}, A.~{Boss}, M.~R. {Haas}, D.~{Buzasi}, and
  D.~{Fischer}.
\newblock {Kepler-16: A Transiting Circumbinary Planet}.
\newblock {\em Science}, 333(6049):1602--1606, September 2011.

\bibitem{2012ApJ...758...87O}
Jerome~A. {Orosz}, William~F. {Welsh}, Joshua~A. {Carter}, Erik {Brugamyer},
  Lars~A. {Buchhave}, William~D. {Cochran}, Michael {Endl}, Eric~B. {Ford},
  Phillip {MacQueen}, Donald~R. {Short}, Guillermo {Torres}, Gur {Windmiller},
  Eric {Agol}, Thomas {Barclay}, Douglas~A. {Caldwell}, Bruce~D. {Clarke},
  Laurance~R. {Doyle}, Daniel~C. {Fabrycky}, John~C. {Geary}, Nader
  {Haghighipour}, Matthew~J. {Holman}, Khadeejah~A. {Ibrahim}, Jon~M.
  {Jenkins}, Karen {Kinemuchi}, Jie {Li}, Jack~J. {Lissauer}, Andrej
  {Pr{\v{s}}a}, Darin {Ragozzine}, Avi {Shporer}, Martin {Still}, and
  Richard~A. {Wade}.
\newblock {The Neptune-sized Circumbinary Planet Kepler-38b}.
\newblock {\em The Astrophysical Journal}, 758(2):87, October 2012.

\bibitem{2012Sci...337.1511O}
Jerome~A. {Orosz}, William~F. {Welsh}, Joshua~A. {Carter}, Daniel~C.
  {Fabrycky}, William~D. {Cochran}, Michael {Endl}, Eric~B. {Ford}, Nader
  {Haghighipour}, Phillip~J. {MacQueen}, Tsevi {Mazeh}, Roberto
  {Sanchis-Ojeda}, Donald~R. {Short}, Guillermo {Torres}, Eric {Agol}, Lars~A.
  {Buchhave}, Laurance~R. {Doyle}, Howard {Isaacson}, Jack~J. {Lissauer},
  Geoffrey~W. {Marcy}, Avi {Shporer}, Gur {Windmiller}, Thomas {Barclay},
  Alan~P. {Boss}, Bruce~D. {Clarke}, Jonathan {Fortney}, John~C. {Geary},
  Matthew~J. {Holman}, Daniel {Huber}, Jon~M. {Jenkins}, Karen {Kinemuchi},
  Ethan {Kruse}, Darin {Ragozzine}, Dimitar {Sasselov}, Martin {Still}, Peter
  {Tenenbaum}, Kamal {Uddin}, Joshua~N. {Winn}, David~G. {Koch}, and William~J.
  {Borucki}.
\newblock {Kepler-47: A Transiting Circumbinary Multiplanet System}.
\newblock {\em Science}, 337(6101):1511, September 2012.

\bibitem{2012ApJ...745L..23Q}
S.~B. {Qian}, L.~Y. {Zhu}, Z.~B. {Dai}, E.~{Fern{\'a}ndez-Laj{\'u}s}, F.~Y.
  {Xiang}, and J.~J. {He}.
\newblock {Circumbinary Planets Orbiting the Rapidly Pulsating Subdwarf B-type
  Binary NY Vir}.
\newblock {\em The Astrophysical Journal Letters}, 745(2):L23, February 2012.

\bibitem{2012MNRAS.422L..24Q}
S.~B. {Qian}, L.~{Liu}, L.~Y. {Zhu}, Z.~B. {Dai}, E.~{Fern{\'a}ndez Laj{\'u}s},
  and G.~L. {Baume}.
\newblock {A circumbinary planet in orbit around the short-period white dwarf
  eclipsing binary RR Cae}.
\newblock {\em Monthly Notices of the Royal Astronomical Society},
  422(1):L24--L27, May 2012.

\bibitem{2012Natur.481..475W}
W.~F. {Welsh}, J.~A. {Orosz}, J.~A. {Carter}, D.~C. {Fabrycky}, E.~B. {Ford},
  J.~J. {Lissauer}, A.~{Pr{\v s}a}, S.~N. {Quinn}, D.~{Ragozzine}, D.~R.
  {Short}, G.~{Torres}, J.~N. {Winn}, L.~R. {Doyle}, T.~{Barclay},
  N.~{Batalha}, S.~{Bloemen}, E.~{Brugamyer}, L.~A. {Buchhave}, C.~{Caldwell},
  D.~A. {Caldwell}, J.~L. {Christiansen}, D.~R. {Ciardi}, W.~D. {Cochran},
  M.~{Endl}, J.~J. {Fortney}, T.~N. {Gautier}, III, R.~L. {Gilliland}, M.~R.
  {Haas}, J.~R. {Hall}, M.~J. {Holman}, A.~W. {Howard}, S.~B. {Howell},
  H.~{Isaacson}, J.~M. {Jenkins}, T.~C. {Klaus}, D.~W. {Latham}, J.~{Li}, G.~W.
  {Marcy}, T.~{Mazeh}, E.~V. {Quintana}, P.~{Robertson}, A.~{Shporer}, J.~H.
  {Steffen}, G.~{Windmiller}, D.~G. {Koch}, and W.~J. {Borucki}.
\newblock {Transiting circumbinary planets Kepler-34 b and Kepler-35 b}.
\newblock {\em Nature}, 481(7382):475--479, January 2012.

\bibitem{2013ApJ...768..127S}
Megan~E. {Schwamb}, Jerome~A. {Orosz}, Joshua~A. {Carter}, William~F. {Welsh},
  Debra~A. {Fischer}, Guillermo {Torres}, Andrew~W. {Howard}, Justin~R.
  {Crepp}, William~C. {Keel}, Chris~J. {Lintott}, Nathan~A. {Kaib}, Dirk
  {Terrell}, Robert {Gagliano}, Kian~J. {Jek}, Michael {Parrish}, Arfon~M.
  {Smith}, Stuart {Lynn}, Robert~J. {Simpson}, Matthew~J. {Giguere}, and Kevin
  {Schawinski}.
\newblock {Planet Hunters: A Transiting Circumbinary Planet in a Quadruple Star
  System}.
\newblock {\em The Astrophysical Journal}, 768(2):127, May 2013.

\bibitem{2014ApJ...781...20K}
Adam~L. {Kraus}, Michael~J. {Ireland}, Lucas~A. {Cieza}, Sasha {Hinkley},
  Trent~J. {Dupuy}, Brendan~P. {Bowler}, and Michael~C. {Liu}.
\newblock {Three Wide Planetary-mass Companions to FW Tau, ROXs 12, and ROXs
  42B}.
\newblock {\em The Astrophysical Journal}, 781(1):20, January 2014.

\bibitem{2014ApJ...784...14K}
V.~B. {Kostov}, P.~R. {McCullough}, J.~A. {Carter}, M.~{Deleuil}, R.~F.
  {D{\'\i}az}, D.~C. {Fabrycky}, G.~{H{\'e}brard}, T.~C. {Hinse}, T.~{Mazeh},
  J.~A. {Orosz}, Z.~I. {Tsvetanov}, and W.~F. {Welsh}.
\newblock {Kepler-413b: A Slightly Misaligned, Neptune-size Transiting
  Circumbinary Planet}.
\newblock {\em The Astrophysical Journal}, 784(1):14, March 2014.

\bibitem{2015ApJ...809...26W}
William~F. {Welsh}, Jerome~A. {Orosz}, Donald~R. {Short}, William~D. {Cochran},
  Michael {Endl}, Erik {Brugamyer}, Nader {Haghighipour}, Lars~A. {Buchhave},
  Laurance~R. {Doyle}, Daniel~C. {Fabrycky}, Tobias~Cornelius {Hinse},
  Stephen~R. {Kane}, Veselin {Kostov}, Tsevi {Mazeh}, Sean~M. {Mills}, Tobias
  W.~A. {M{\"u}ller}, Billy {Quarles}, Samuel~N. {Quinn}, Darin {Ragozzine},
  Avi {Shporer}, Jason~H. {Steffen}, Lev {Tal-Or}, Guillermo {Torres}, Gur
  {Windmiller}, and William~J. {Borucki}.
\newblock {Kepler 453 b - The 10th Kepler Transiting Circumbinary Planet}.
\newblock {\em The Astrophysical Journal}, 809(1):26, August 2015.

\bibitem{2016AJ....152..125B}
D.~P. {Bennett}, S.~H. {Rhie}, A.~{Udalski}, A.~{Gould}, Y.~{Tsapras},
  D.~{Kubas}, I.~A. {Bond}, J.~{Greenhill}, A.~{Cassan}, N.~J. {Rattenbury},
  T.~S. {Boyajian}, J.~{Luhn}, M.~T. {Penny}, J.~{Anderson}, F.~{Abe},
  A.~{Bhattacharya}, C.~S. {Botzler}, M.~{Donachie}, M.~{Freeman}, A.~{Fukui},
  Y.~{Hirao}, Y.~{Itow}, N.~{Koshimoto}, M.~C.~A. {Li}, C.~H. {Ling},
  K.~{Masuda}, Y.~{Matsubara}, Y.~{Muraki}, M.~{Nagakane}, K.~{Ohnishi},
  H.~{Oyokawa}, Y.~C. {Perrott}, To. {Saito}, A.~{Sharan}, D.~J. {Sullivan},
  T.~{Sumi}, D.~{Suzuki}, P.~J. {Tristram}, A.~{Yonehara}, P.~C.~M. {Yock},
  {MOA Collaboration}, M.~K. {Szyma{\'n}ski}, I.~{Soszy{\'n}ski}, K.~{Ulaczyk},
  {\L}.~{Wyrzykowski}, {OGLE Collaboration}, W.~{Allen}, D.~{DePoy},
  A.~{Gal-Yam}, B.~S. {Gaudi}, C.~{Han}, I.~A.~G. {Monard}, E.~{Ofek}, R.~W.
  {Pogge}, {{\ensuremath{\mu}}FUN Collaboration}, R.~A. {Street}, D.~M.
  {Bramich}, M.~{Dominik}, K.~{Horne}, C.~{Snodgrass}, I.~A. {Steele}, {Robonet
  Collaboration}, M.~D. {Albrow}, E.~{Bachelet}, V.~{Batista}, J.~P.
  {Beaulieu}, S.~{Brillant}, J.~A.~R. {Caldwell}, A.~{Cole}, C.~{Coutures},
  S.~{Dieters}, D.~{Dominis Prester}, J.~{Donatowicz}, P.~{Fouqu{\'e}},
  M.~{Hundertmark}, U.~G. {J{\o}rgensen}, N.~{Kains}, S.~R. {Kane}, J.~B.
  {Marquette}, J.~{Menzies}, K.~R. {Pollard}, C.~{Ranc}, K.~C. {Sahu},
  J.~{Wambsganss}, A.~{Williams}, M.~{Zub}, and {PLANET Collaboration}.
\newblock {The First Circumbinary Planet Found by Microlensing:
  OGLE-2007-BLG-349L(AB)c}.
\newblock {\em The Astronomical Journal}, 152(5):125, November 2016.

\bibitem{2016ApJ...827...86K}
Veselin~B. {Kostov}, Jerome~A. {Orosz}, William~F. {Welsh}, Laurance~R.
  {Doyle}, Daniel~C. {Fabrycky}, Nader {Haghighipour}, Billy {Quarles},
  Donald~R. {Short}, William~D. {Cochran}, Michael {Endl}, Eric~B. {Ford}, Joao
  {Gregorio}, Tobias~C. {Hinse}, Howard {Isaacson}, Jon~M. {Jenkins}, Eric
  L.~N. {Jensen}, Stephen {Kane}, Ilya {Kull}, David~W. {Latham}, Jack~J.
  {Lissauer}, Geoffrey~W. {Marcy}, Tsevi {Mazeh}, Tobias W.~A. {M{\"u}ller},
  Joshua {Pepper}, Samuel~N. {Quinn}, Darin {Ragozzine}, Avi {Shporer},
  Jason~H. {Steffen}, Guillermo {Torres}, Gur {Windmiller}, and William~J.
  {Borucki}.
\newblock {Kepler-1647b: The Largest and Longest-period Kepler Transiting
  Circumbinary Planet}.
\newblock {\em The Astrophysical Journal}, 827(1):86, August 2016.

\bibitem{2017MNRAS.468.2932G}
A.~K. {Getley}, B.~{Carter}, R.~{King}, and S.~{O'Toole}.
\newblock {Evidence for a planetary mass third body orbiting the binary star
  KIC 5095269}.
\newblock {\em Monthly Notices of the Royal Astronomical Society},
  468(3):2932--2937, July 2017.

\bibitem{2017MNRAS.468L.118J}
Chetana {Jain}, Biswajit {Paul}, Rahul {Sharma}, Abdul {Jaleel}, and Anjan
  {Dutta}.
\newblock {Indication of a massive circumbinary planet orbiting the low-mass
  X-ray binary MXB 1658-298}.
\newblock {\em Monthly Notices of the Royal Astronomical Society},
  468(1):L118--L122, June 2017.

\bibitem{2018A&A...619A..43A}
R.~{Asensio-Torres}, M.~{Janson}, M.~{Bonavita}, S.~{Desidera}, C.~{Thalmann},
  M.~{Kuzuhara}, Th. {Henning}, F.~{Marzari}, M.~R. {Meyer},
  P.~{Calissendorff}, and T.~{Uyama}.
\newblock {SPOTS: The Search for Planets Orbiting Two Stars. III. Complete
  sample and statistical analysis}.
\newblock {\em Astronomy and Astrophysics}, 619:A43, October 2018.

\bibitem{2021AJ....162..234K}
Veselin~B. {Kostov}, Brian~P. {Powell}, Jerome~A. {Orosz}, William~F. {Welsh},
  William {Cochran}, Karen~A. {Collins}, Michael {Endl}, Coel {Hellier},
  David~W. {Latham}, Phillip {MacQueen}, Joshua {Pepper}, Billy {Quarles},
  Lalitha {Sairam}, Guillermo {Torres}, Robert~F. {Wilson}, Serge {Bergeron},
  Pat {Boyce}, Allyson {Bieryla}, Robert {Buchheim}, Caleb {Ben Christiansen},
  David~R. {Ciardi}, Kevin~I. {Collins}, Dennis~M. {Conti}, Scott {Dixon}, Pere
  {Guerra}, Nader {Haghighipour}, Jeffrey {Herman}, Eric~G. {Hintz}, Ward~S.
  {Howard}, Eric L.~N. {Jensen}, John~F. {Kielkopf}, Ethan {Kruse}, Nicholas~M.
  {Law}, David {Martin}, Pierre F.~L. {Maxted}, Benjamin~T. {Montet}, Felipe
  {Murgas}, Matt {Nelson}, Greg {Olmschenk}, Sebastian {Otero}, Robert
  {Quimby}, Michael {Richmond}, Richard~P. {Schwarz}, Avi {Shporer}, Keivan~G.
  {Stassun}, Denise~C. {Stephens}, Amaury H.~M.~J. {Triaud}, Joe {Ulowetz},
  Bradley~S. {Walter}, Edward {Wiley}, David {Wood}, Mitchell {Yenawine}, Eric
  {Agol}, Thomas {Barclay}, Thomas~G. {Beatty}, Isabelle {Boisse}, Douglas~A.
  {Caldwell}, Jessie {Christiansen}, Knicole~D. {Col{\'o}n}, Magali {Deleuil},
  Laurance {Doyle}, Michael {Fausnaugh}, G{\'a}bor {F{\H{u}}r{\'e}sz}, Emily~A.
  {Gilbert}, Guillaume {H{\'e}brard}, David~J. {James}, Jon {Jenkins},
  Stephen~R. {Kane}, Jr. {Kidwell}, Richard~C., Ravi {Kopparapu}, Gongjie {Li},
  Jack~J. {Lissauer}, Michael~B. {Lund}, Steve~R. {Majewski}, Tsevi {Mazeh},
  Samuel~N. {Quinn}, Elisa {Quintana}, George {Ricker}, Joseph~E. {Rodriguez},
  Jason {Rowe}, Alexander {Santerne}, Joshua {Schlieder}, Sara {Seager},
  Matthew~R. {Standing}, Daniel~J. {Stevens}, Eric~B. {Ting}, Roland
  {Vanderspek}, and Joshua~N. {Winn}.
\newblock {TIC 172900988: A Transiting Circumbinary Planet Detected in One
  Sector of TESS Data}.
\newblock {\em The Astronomical Journal}, 162(6):234, December 2021.

\bibitem{1975ApJ...195L..65B}
James~M. {Bardeen} and Jacobus~A. {Petterson}.
\newblock {The Lense-Thirring Effect and Accretion Disks around Kerr Black
  Holes}.
\newblock {\em The Astrophysical Journal}, 195:L65--L67, January 1975.

\bibitem{1998ApJ...492L..59S}
Luigi {Stella} and Mario {Vietri}.
\newblock {Lense-Thirring Precession and Quasi-periodic Oscillations in
  Low-Mass X-Ray Binaries}.
\newblock {\em The Astrophysical Journal Letters}, 492(1):L59--L62, January
  1998.

\bibitem{Pen2002}
R.~Penrose.
\newblock \virg{Golden Oldie}: {Gravitational} {Collapse}: {The} {Role} of
  {General} {Relativity}.
\newblock {\em General Relativity and Gravitation}, 34(7):1141--1165, July
  2002.

\bibitem{2009SSRv..148...37S}
Gerhard {Sch{\"a}fer}.
\newblock {Gravitomagnetism in Physics and Astrophysics}.
\newblock {\em Space Science Reviews}, 148(1-4):37--52, December 2009.

\bibitem{StellaPossenti09}
L.~{Stella} and A.~{Possenti}.
\newblock {Lense-Thirring Precession in the Astrophysical Context}.
\newblock {\em Space Science Reviews}, 148(1-4):105--121, December 2009.

\bibitem{2014IAUS..293..125W}
William~F. {Welsh}, Jerome~A. {Orosz}, Joshua~A. {Carter}, and Daniel~C.
  {Fabrycky}.
\newblock {Recent Kepler Results On Circumbinary Planets}.
\newblock In Nader {Haghighipour}, editor, {\em Formation, Detection, and
  Characterization of Extrasolar Habitable Planets}, volume 293, pages
  125--132. Cambridge University Press, Cambridge, April 2014.

\bibitem{2017PhRvD..95h4024C}
Chandrachur {Chakraborty}, Prashant {Kocherlakota}, Mandar {Patil}, Sudip
  {Bhattacharyya}, Pankaj~S. {Joshi}, and Andrzej {Kr{\'o}lak}.
\newblock {Distinguishing Kerr naked singularities and black holes using the
  spin precession of a test gyro in strong gravitational fields}.
\newblock {\em Physical Review D}, 95(8):084024, April 2017.

\bibitem{2018PhRvD..98d3021C}
Chandrachur {Chakraborty} and Sudip {Bhattacharyya}.
\newblock {Does the gravitomagnetic monopole exist? A clue from a black hole
  x-ray binary}.
\newblock {\em Physical Review D}, 98(4):043021, August 2018.

\bibitem{1998MNRAS.297L..76P}
F.~P. {Pijpers}.
\newblock {Helioseismic determination of the solar gravitational quadrupole
  moment}.
\newblock {\em Monthly Notices of the Royal Astronomical Society},
  297(3):L76--L80, July 1998.

\bibitem{2006NewA...11..490I}
L.~{Iorio}.
\newblock {Are we far from testing general relativity with the transitting
  extrasolar planet HD 209458b \virg{Osiris}?}
\newblock {\em New Astronomy}, 11(7):490--494, May 2006.

\bibitem{2008ApJ...685..543J}
A.~{Jord{\'a}n} and G.~{\'A}. {Bakos}.
\newblock {Observability of the General Relativistic Precession of Periastra in
  Exoplanets}.
\newblock {\em The Astrophysical Journal}, 685(1):543--552, September 2008.

\bibitem{2008MNRAS.389..191P}
A.~{P{\'a}l} and B.~{Kocsis}.
\newblock {Periastron precession measurements in transiting extrasolar
  planetary systems at the level of general relativity}.
\newblock {\em Monthly Notices of the Royal Astronomical Society},
  389(1):191--198, September 2008.

\bibitem{2009IAUS..253..492J}
A.~{Jord{\'a}n} and G.~{\'A}. {Bakos}.
\newblock {Observability of the General Relativistic Precession of Periastra in
  Exoplanets}.
\newblock In {\em IAU Symposium}, volume 253 of {\em IAU Symposium}, pages
  492--495, February 2009.

\bibitem{2009ApJ...698.1778R}
D.~{Ragozzine} and A.~S. {Wolf}.
\newblock {Probing the Interiors of very Hot Jupiters Using Transit Light
  Curves}.
\newblock {\em The Astrophysical Journal}, 698(2):1778--1794, June 2009.

\bibitem{2011A&A...535A.116D}
C.~{Damiani} and A.~F. {Lanza}.
\newblock {Prospecting transit duration variations in extrasolar planetary
  systems}.
\newblock {\em Astronomy \& Astrophysics}, 535:A116, November 2011.

\bibitem{2011PASJ...63..287F}
A.~{Fukui}, N.~{Narita}, P.~J. {Tristram}, T.~{Sumi}, F.~{Abe}, Y.~{Itow},
  D.~J. {Sullivan}, I.~A. {Bond}, T.~{Hirano}, M.~{Tamura}, D.~P. {Bennett},
  K.~{Furusawa}, F.~{Hayashi}, J.~B. {Hearnshaw}, S.~{Hosaka}, K.~{Kamiya},
  S.~{Kobara}, A.~{Korpela}, P.~M. {Kilmartin}, W.~{Lin}, C.~H. {Ling},
  S.~{Makita}, K.~{Masuda}, Y.~{Matsubara}, N.~{Miyake}, Y.~{Muraki},
  M.~{Nagaya}, K.~{Nishimoto}, K.~{Ohnishi}, K.~{Omori}, Y.~{Perrott},
  N.~{Rattenbury}, T.~{Saito}, L.~{Skuljan}, D.~{Suzuki}, W.~L. {Sweatman}, and
  K.~{Wada}.
\newblock {Measurements of Transit Timing Variations for WASP-5b}.
\newblock {\em Publications of the Astronomical Society of Japan},
  63(1):287--300, February 2011.

\bibitem{2011Ap&SS.331..485I}
L.~{Iorio}.
\newblock {Classical and relativistic node precessional effects in WASP-33b and
  perspectives for detecting them}.
\newblock {\em Astrophysics and Space Science}, 331(2):485--496, February 2011.

\bibitem{2011MNRAS.411..167I}
L.~{Iorio}.
\newblock {Classical and relativistic long-term time variations of some
  observables for transiting exoplanets}.
\newblock {\em Monthly Notices of the Royal Astronomical Society},
  411(1):167--183, February 2011.

\bibitem{2012MNRAS.423.1381E}
M.~T. {Eibe}, L.~{Cuesta}, A.~{Ull{\'a}n}, A.~{P{\'e}rez-Verde}, and
  J.~{Navas}.
\newblock {Analysis of variations in transit time and transit duration in
  WASP-3. Evidence of secular perturbations reconsidered}.
\newblock {\em Monthly Notices of the Royal Astronomical Society},
  423(2):1381--1389, June 2012.

\bibitem{2012ApJ...757..105K}
Stephen~R. {Kane}, Jonathan {Horner}, and Kaspar {von Braun}.
\newblock {Cyclic Transit Probabilities of Long-period Eccentric Planets due to
  Periastron Precession}.
\newblock {\em The Astrophysical Journal}, 757(1):105, September 2012.

\bibitem{2012Ap&SS.341..323L}
Lin-Sen {Li}.
\newblock {Parameterized post-Newtonian orbital effects in extrasolar planets}.
\newblock {\em Astrophysics and Space Science}, 341(2):323--330, October 2012.

\bibitem{2013RAA....13.1231Z}
Shan-Shan {Zhao} and Yi~{Xie}.
\newblock {Parametrized post-Newtonian secular transit timing variations for
  exoplanets}.
\newblock {\em Research in Astronomy and Astrophysics}, 13(10):1231--1239,
  October 2013.

\bibitem{2019A&A...628A..80B}
Luc {Blanchet}, Guillaume {H{\'e}brard}, and Fran{\c{c}}ois {Larrouturou}.
\newblock {Detecting the general relativistic orbital precession of the
  exoplanet HD 80606b}.
\newblock {\em Astronomy \& Astrophysics}, 628:A80, August 2019.

\bibitem{2021MNRAS.505.1567A}
G.~{Antoniciello}, L.~{Borsato}, G.~{Lacedelli}, V.~{Nascimbeni},
  O.~{Barrag{\'a}n}, and R.~{Claudi}.
\newblock {Detecting general relativistic orbital precession in transiting hot
  Jupiters}.
\newblock {\em Monthly Notices of the Royal Astronomical Society},
  505(2):1567--1574, August 2021.

\bibitem{2021E&ES..658a2051G}
Xirui {Gou}, Xinyue {Pan}, and Le~{Wang}.
\newblock {General Relativity Testing in Exoplanetary Systems}.
\newblock In {\em IOP Conference Series: Earth and Environmental Science},
  volume 658 of {\em IOP Conference Series: Earth and Environmental Science},
  page 012051, February 2021.

\end{thebibliography}

\end{document}